\documentclass[final,1p,times]{elsarticle} 
\usepackage{graphicx} 
\usepackage{amssymb} 
\usepackage{amsthm} 
\usepackage{lineno} 

\usepackage{amsmath}
\newcommand{\beq}{\begin{equation}}
\newcommand{\eeq}{\end{equation}}
\newcommand{\beqa}{\begin{eqnarray}}
\newcommand{\eeqa}{\end{eqnarray}}
\def\r{{\boldsymbol r}}
\def\z{{\boldsymbol z}}
\def\x{{\boldsymbol x}}

\def\k{{\boldsymbol k}}
\def\q{{\boldsymbol q}}

\def\0{{\boldsymbol 0}}

\def\cal{\mathcal}

\journal{Nuclear Physics A} 
\begin{document} 

\begin{frontmatter} 


\title{Heavy-quarks in the QGP: study of medium effects through euclidean propagators and spectral functions}

\author{A. Beraudo}

\address{Dipartimento di Fisica Teorica dell'Universit\`a  di Torino and INFN, sezione di Torino}

\author{J.P. Blaizot}

\address{IPhT, CEA-Saclay}



\author{G. Garberoglio$^\dagger$ and P. Faccioli$^\star$}

\address{Dipartimento di Fisica dell'Universit\`a  di Trento, $^\dagger$CNISM and $^\star$INFN, gruppo collegato di Trento}

\begin{abstract} 
The heavy-quark spectral function in a hot plasma is reconstructed from the corresponding euclidean propagator. The latter is evaluated through a path-integral simulation. A weak-coupling calculation is also performed, allowing to interpret the qualitative behavior of the spectral function in terms of quite general physical processes. 

\end{abstract} 

\end{frontmatter} 



\section{Introduction}\label{sec:intro}
Heavy Quarks (HQs), being produced in the very early stages of nucleus-nucleus collisions, have long been used as probes of the resulting hot (possibly deconfined) medium.
Medium modifications of heavy-quark spectra (elliptic flow, suppression of high-$p_T$ particles) can shed light on the degree of their interaction and thermalization with the rest of the plasma.
As a result, it is interesting to perform a first-principle calculation of the heavy-quark spectral function in a hot plasma, showing how for the latter an interesting qualitative structure arises from the interaction with the medium.
We first perform a Hard Thermal Loop (HTL) weak-coupling calculation, allowing to identify quite general physical processes leading to a modification of the vacuum result. We then formulate the problem in terms of a path-integral which is evaluated exactly, with a
Monte Carlo algorithm.  From such  an euclidean correlator, we can
shed light on the structure of the corresponding non-perturbative
in-medium spectral density.

\section{HQ spectral function: weak-coupling calculation}\label{sec:weak}
The in-medium analytic propagator of a non-relativistic heavy-quark reads
\begin{equation}
G(z,p)=\frac{-1}{z-E_p-\Sigma(z,p)},
\end{equation}
where $E_p\!=\!M+p^2/2M$, $\,\Sigma$ is the HQ self-energy and setting $z\!=\!\omega+i\,\eta$ gives the \emph{retarded propagator} $G^R(\omega)\!=\!G(\omega\!+\!i\,\eta)$. The imaginary part of the latter provides the HQ spectral function
\beq
\sigma(\omega)\equiv 2{\rm Im}\,G^R(\omega)=\frac{\Gamma(\omega)}{[\omega-E_p-{\rm Re}\,\Sigma(\omega)]^2+\Gamma^2(\omega)/4},
\eeq
where $\Gamma(\omega)\!\equiv\!-2{\rm Im}\,\Sigma^R(\omega)$. The HQ spectral function is then non-vanishing only for energies for which the self-energy develops an imaginary-part. For a non-relativistic heavy quark at zero momentum (playing the role of an external probe) placed in the QGP the self-energy can be evaluated in the (HTL-resummed) one-loop approximation
\beq
\Sigma(p^0)\!=\!g^2 C_F\!\int\frac{d\k}{(2\pi)^3}\!\int_{0}^{+\infty}\frac{dk^0}{2\pi}\rho_L(k^0,k)\left[\frac{1+N(k^0)}{p^0-(E_k+k^0)}+
\frac{N(k^0)}{p^0-(E_k-k^0)}\right],
\eeq
which is expressed in terms of the HTL spectral function of a longitudinal gluon \cite{leb}
\beq\label{eq:HTL_long_spec}
\rho_L(\omega >0,q)\equiv 2\pi\,\left[Z_L(q)\,\delta(\omega\!-\!\omega_L(q))
+\theta(q^2\!-\omega^2)\,\beta_L(\omega,q)\right].
\eeq
The latter displays a peak for time-like momenta, arising from the propagation of a \emph{plasma wave}, and a continuum part for space-like momenta, describing the \emph{Landau damping} (i.e. soft collisions with the plasma particle).
\begin{figure}[!tp]
\begin{center}
\includegraphics[clip,width=0.6\textwidth]{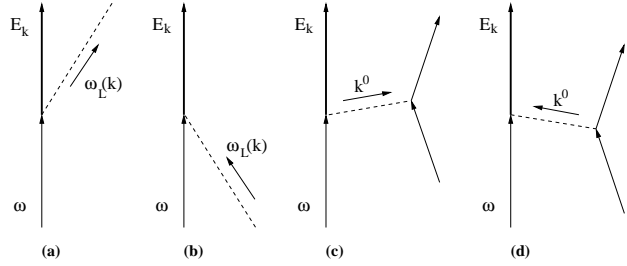}
\caption{The different processes contributing to the imaginary part of the HQ self-energy.}
\label{fig:processes} 
\end{center}
\end{figure}
The imaginary part of the HQ self-energy gets then a pole and a continuum contribution from the different processes shown in Fig. \ref{fig:processes}. The HQ can go on-shell through the emission/absorption (with obvious meaning of the statistical factors) of a plasmon
\begin{multline}
\Gamma^{\rm pole}(\omega)=g^2 C_F\int\frac{d\k}{(2\pi)^3}\,(2\pi)\,Z_L(k)\times\\
\times\big\{\left(1+N(\omega_L(k))\right)\delta\left[\omega-(E_k+\omega_L(k))\right]+N(\omega_L(k))\,\delta\left[\omega-(E_k-\omega_L(k))\right]\big\}
\end{multline}
or of a space-like (resummed) gluon exchanged with the medium particles
\begin{multline}
\Gamma^{\rm cont}(\omega)=g^2 C_F\int\frac{d\k}{(2\pi)^3}\int_{0}^{k}dk^0\,\beta_L(k^0,k)\,\times\\
\times(2\pi)\left\{\left[1+N(k^0)\right]\delta\left[\omega-(E_k+k^0)\right]+N(k^0)\,\delta\left[\omega-(E_k-k^0)\right]\right\}.
\end{multline}
\begin{figure}[!tp]
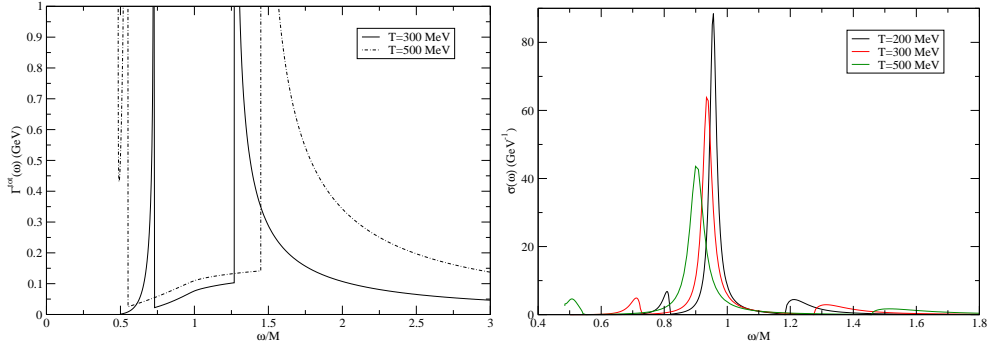

\begin{center}
\includegraphics[clip,width=0.475\textwidth]{gamma_tot_def.eps}
\includegraphics[clip,width=0.475\textwidth]{spectral.eps}
\caption{The imaginary part of the HQ self-energy in the QGP and the resulting spectral function for different temperatures, setting $M\!=\!1.5$ GeV and $\,\alpha_s\!=\!0.3$.}
\label{fig:spectral} 
\end{center}
\end{figure}
The results for the corresponding in-medium spectral function are displayed in Fig. \ref{fig:spectral}. Its main features are: a \emph{broadening} and a \emph{negative shift} of the main peak and the appearance of \emph{secondary peaks} at energies corresponding to a huge density of states for the absorption/emission of a plasmon. The bumps arise from divergences in $\Gamma^{\rm pole}(\omega)$ (\emph{Van-Hove singularities}).  

\section{HQ spectral function: path-integral approach}\label{sec:path}
In order to formulate the problem in terms of a path-integral we employ the following strategy, borrowed from \cite{bbr1,bbr2}, and extended to the case of finite-mass quarks. 
We consider the propagation of the HQ in the background of a gauge field, summing over all its possible trajectories and integrating over the field  configurations with an action accounting for the presence of an hot medium.
The field configurations are thus weighted according to the HTL effective action, which allows to properly dress the propagation of soft (long- wavelength) modes with the interaction with the plasma particles.
Neglecting non-abelian effects one takes the above action to be gaussian and this allows to perform the functional integral over the field configurations exactly, thus reducing by orders of magnitude the required numerical efforts with respect to a lQCD calculation. In practice we are addressing the simpler case of a hot-QED plasma, for which the above procedure is exact, fixing at the end the parameters entering in the calculation (temperature, coupling and HQ mass) to values of phenomenological interest for the study of the QGP. We make the choice of losing some peculiar feature of QCD, to investigate \emph{exactly} (without any $T/M$ expansion) and with a huge statistics (not conceivable in a lQCD simulation) important and very general medium effects. 

\begin{figure}[!tp]
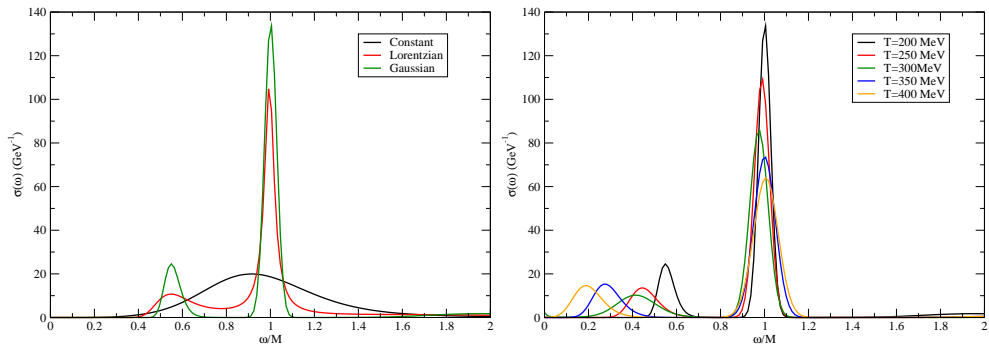

\begin{center}
\includegraphics[clip,width=0.475\textwidth]{MEM_T1_new.eps}
\includegraphics[clip,width=0.475\textwidth]{MEM_Tscan.eps}
\caption{The HQ spectral density from the MEM procedure, for various choices of the default model at $T\!=\!200$ MeV (left) and for different temperatures with a gaussian default (right). We set $M\!=\!1.5$ GeV and $\,\alpha_s\!=\!0.3$.}
\label{fig:mem} 
\end{center}
\end{figure}

We thus perform numerical simulations for the following euclidean correlator, describing the propagation of a HQ from $(0,\0)$ to $(\tau,\r)$,
\begin{multline}
G(\tau,\r)
=\int_{\z(0)=\0'}^{\z(\tau)=\r}[{\cal D}
\z]\exp\left[-\int_{0}^{\tau}d\tau'\left(M+\frac{1}{2}M\dot
{\z}^2\right)\right]\times\\
\times
\exp\left[\frac{g^2}{2}\int_0^\tau d\tau' \int_0^\tau d\tau''
\Delta_{L}^T(\tau'-\tau'',\z(\tau')-\z(\tau''))\right],
\end{multline}
which results expressed in terms of the HTL longitudinal gauge-boson propagator (after subtraction of the instantaneous vacuum Coulomb interaction)
\beq
\Delta_{L}(\tau,\q)=\int_{-\infty}^{+\infty}\frac{dq_0}{2\pi}e^{-q_0\tau}\rho_L(q_0,\q)
[\theta(\tau)+N(q^0)],
\eeq
with $\rho_L$ given in Eq. (\ref{eq:HTL_long_spec}). By integrating over all possible final positions we project it to zero momentum and reconstruct the spectral function of a HQ at rest in the hot plasma, after inverting the relation
\beq
G(\tau,p=0)=\int\frac{d\omega}{2\pi}\,e^{-\omega\tau}\sigma(\omega,p=0).
\eeq
For this we employ a Maximum Entropy Method \cite{mem} algorithm. The latter allows to find the most probable spectral density compatible with the data and some prior knowledge embodied in a default model.
In Fig. \ref{fig:mem} we display the results obtained with different choices of the default model and for various temperatures. With a flat default model one gets simply a very broad bump, with a peak slightly shifted to lower energies.
On the contrary, inserting richer information into the prior knowledge -- e.g. the width and mass shift arising from the static ($M\!=\!\infty$) case and the request of fulfilling a few sum rules -- the spectral density displays a qualitative behavior (with small bumps away from the main peak) reminiscent of the one found in the perturbative calculation. Remarkably, these findings arise from the inversion algorithm itself, due to the high quality of the data, without putting any information on the presence of secondary peaks in the default model. In particular the shift of the low-energy bump with the temperature appears consistent with its physical interpretation in terms of plasmon-absorption.
\section{Conclusions}\label{sec:concl}
We displayed results for the HQ spectral function in a hot plasma, like the QGP. A HTL weak-coupling calculation allowed to identify important qualitative features of the latter (broadening and shift of the main peak and appearance of secondary bumps), relating them to well-defined physical processes.
The problem was then addressed  non-perturbatively, through a
path-integral simulation. Also in this case, a non-trivial
structure was found in the spectrum.
Further details will be given elsewhere \cite{bbfg}.
It will be of interest to compare our findings with lQCD results \cite{kar}.



\end{document}